\documentclass[12pt]{article}

\usepackage{amsmath}
\usepackage{amsfonts}
\usepackage{amssymb}
\usepackage[numbers,sort&compress]{natbib}

\def\be{\begin{equation}}
\def\ee{\end{equation}}
\def\bga{\begin{gather}}
\def\ega{\end{gather}}

\def\lsim{\mathrel{\rlap{\lower3pt\hbox{\hskip0pt$\sim$}}
     \raise1pt\hbox{$<$}}}         
\def\gsim{\mathrel{\rlap{\lower3pt\hbox{\hskip0pt$\sim$}}
     \raise1pt\hbox{$>$}}}         

\newcommand{\comment}[1]{}

\begin{document}

\numberwithin{equation}{section}

\begin{flushright}

{NYU-TH-10/12/2016}
\end{flushright}

\vskip 1cm

\begin{center}

{\Large {Dark Energy and Inflation with Volume Normalized Action}}

\end{center}

\thispagestyle{empty}

\vskip 0.5 cm

\begin{center}

{\large  {Gregory Gabadadze and Siqing Yu}}

 \vspace{.2cm}

{\it Center for Cosmology and Particle Physics,
Department of Physics,}
\centerline{\it New York University, New York,
NY, 10003, USA}

\end{center}

\vskip 1cm

\begin{abstract}
In the theories described by a volume normalized action functional, an arbitrary cosmological constant is eliminated from the physical picture of our Universe, and dynamical alternatives must be responsible  for today's accelerated expansion as well as the conjectured early-time inflation in the Universe.  A few well-known such scenarios realized by a single homogeneous scalar field are examined in this new context,  and their diverse fates are elucidated.
Typical inflationary models are not affected at the level of the background evolution,  and also give rise
to the scalar perturbations equivalent to those obtained in the standard General Relativity; however,
the primordial quantum tensor fluctuations are absent in the new framework,  irrespective of the inflationary model.
As a consequence,  our proposal would  be ruled out should the primordial tensor modes, or their indirect 
consequences,  be observed experimentally. 
\end{abstract}

\newpage

\section{Introduction}\label{intro}
Based on the ideas in \cite{Linde:1988ws,Tseytlin:1990hn}, a class of theories that eliminate an arbitrary cosmological constant (CC) from our Universe has been proposed recently \cite{Gabadadze:2014rwa, ggsy}. It is suggested that the 4D physical metric of our Universe, $g_{\mu \nu}$,  lives on the conformal boundary of a 
5D anti-de Sitter (AdS$_5$) space, a bulk space that is endowed with its own 5D metric $F_{AB}$. The theories are described by the action functionals of the form
\begin{equation}\label{pr1}
\mathcal{A}=\frac{V_F}{V_g}S+S_F,
\end{equation}
where $V_F$ and $V_g$ are the respective invariant volumes of the 5D  and 4D spacetimes, 
$S_F$ is the bulk Einstein-Hilbert action with a large negative CC, and $S$ is the action of our world, 
containing $g$-gravity and matter fields. The full action (\ref{pr1}) extends the effective action that describes the classical evolution of our Universe introduced by Tseytlin \cite {Tseytlin:1990hn},
\begin{equation}\label{ac2}
\bar{S}=\frac{S}{V_g}.
\end{equation}
We hereafter call $\bar{S}$ the ``volume normalized" action, a term inspired by the works 
\cite{Davidson, Davidson2}, in which Davidson and Rubin have named the theory described by $S_{\rm NGR}=\bar{S}/\epsilon$ ``normalized General Relativity," with $\epsilon$ being an unspecified parameter having mass dimension 4.

Both (\ref{pr1}) and (\ref{ac2}) have the property that any CC generated in the 4D $g$-universe by any source is immaterial for the $g$-universe itself. For (\ref{pr1}), that CC gets absorbed into $S_F$, adding to or subtracting from the curvature of the bulk AdS$_5$ universe to give it a small correction. A high price for this resolution of the CC problem is an intrinsic spacetime nonlocality, which however is only sensed by CC, but not any other form of energy/matter \cite{Gabadadze:2014rwa, ggsy}. 

The full action (\ref{pr1}) has been introduced for its theoretical advantage of producing a consistent low energy theory of quantum gravity in the $g$-universe. As noted by Tseytlin \cite {Tseytlin:1990hn}, with just (\ref{ac2}), the $g$ gravity loops should be calculated with an effective Planck constant $\hbar V_g$, leading to large quantum gravity effects which would at least ruin the classical solution that eliminates the CC.  On the contrary, using (\ref{pr1}), the particle physics quantum effects can be kept intact, while the gravity 
loops of $g$ are suppressed by a different effective Planck constant 
$\hbar V_g/V_F$. The ratio of the volumes is determined by the classical solutions such that $\hbar V_g/V_F\to 0$. Then the gravity loops do not generate 
any counterterms  to the  action  for the metric $g$, and 
at low energies, gravity in our Universe becomes a purely classical theory that  couples to quantized particle 
physics \cite{Gabadadze:2014rwa,ggsy}. Therefore, when we talk about classical cosmologies in the $g$-universe, we can safely use (\ref{ac2}), but we really need to bear in mind the full action (\ref{pr1}) when dealing with quantum effects.

This CC-free framework then necessitates a dynamical dark energy component that drives the accelerated expansion today. In this regard, \cite{ggsy2} utilizes massive gravity \cite{mgr1,mgr2}, which is ghost-free \cite{ghostfree1, ghostfree2, ghostfree3, gfree4} and has a technically natural graviton mass \cite{dRGHP}; it also presents  
a new class of healthy self-accelerated background solutions in the decoupling limit. With the help of nonlocal Galileon terms \cite{galileon}, these solutions exhibit no linear scalar-tensor mixing \cite {dRGHP_sa} and thus no van Dam-Veltman-Zakharov (vDVZ) discontinuity \cite{vDVZ1,vDVZ2}.  However, just as any other solution obtained via the decouplimg limit, their full-theory consistency beyond that limit remains unsettled. Reviews of massive gravity can be found in \cite{mgreview1,mgreview2,mgreview3}.

Here we study instead some ``more traditional" dark energy candidates in the new framework. We limit ourselves to the following possibilities described by the rolling of a single homogeneous scalar field: (a) minimally coupled to gravity, commonly known as ``quintessence" \cite{quint11,quint22}, (b) non-minimally coupled to gravity \cite{nmcintro,nmcintro2}, and (c) Chaplygin gas \cite{chap}. Furthermore, many of the single scalar field scenarios with a slow-roll regime were first used to describe inflation \cite{inf,inf2,inf3}, for which the results of this paper apply equally well, once the energy scales involved are adjusted accordingly.\footnote{Throughout the
paper, we consider only the scenarios in which our 4D Universe evolves in
a single vacuum state. We do not attempt, for instance, to generalize (\ref{ac2}) to adopt chaotic inflation
with multiple vacua.} Conveniently, we will treat the dark energy and the inflationary scenarios driven by the same mechanism alike, commenting on them separately when appropriate. It turns out that (a) and (b) are unscathed when incorporated into (\ref{ac2}), whereas (c) loses its original features. The details are given in sections \ref{mcs}, \ref{nmcs}, and \ref{cg}, respectively. We then conclude with a discussion on the classical and quantum field fluctuations in section \ref{conc}, highlighting the testable prediction of a zero tensor-to-scalar ratio in the primordial power spectra. Reviews can be found in \cite{dereview1,dereview2,dereview3} for dark energy and \cite{infreview1,infreview2,infreview3,infreview4,infreview5} for inflation.

To discuss cosmology, we use a homogeneous scalar profile, $\phi(\vec{x},t)=\phi(t)$, on a flat Friedmann-Robertson-Walker (FRW) background metric in the mostly positive signature, $ds^2=-dt^2+a(t)^2\delta_{ij}dx^idx^j$. Time derivatives are denoted by dots. To make nonlocal expressions unambiguous, a ratio between spacetime integrals is defined as follows: first compute the ratio of integrals in a finite spacetime volume, then take the limit when the region of integration tend to infinity in all dimensions.

We note that the works \cite{Davidson, Davidson2} by Davidson and Rubin have discussed extensively the classical FRW cosmologies allowed by the effective action (\ref{ac2}). Using mostly minisuperspace techniques, \cite{Davidson} has showed the vanishing of a rigid CC in quintessential models, and \cite{Davidson2} has proved that a perfect fluid can only lead to nonclosed universes and a zero CC, except for the exotic cases of closed static Einstein universe and Eddington-Lema\^itre universe \cite{el1,el2,el3}. As such, we acknowledge that the classical analysis in sections \ref{mcs}, \ref{nmcs}, and \ref{cg} have significant conceptual overlaps with \cite{Davidson, Davidson2}. We also note that our major difference from \cite{Davidson, Davidson2} is to establish the consistent low energy theory of quantum gravity through the full action (\ref{pr1}), a problem left unaddressed in \cite{Davidson, Davidson2}. 

\section{Minimally Coupled Scalar}\label{mcs}
In the framework of (\ref{ac2}), the theory of a scalar field minimally coupled to gravity is described by the variation of
\begin{equation}
\bar{S}=\frac{S}{V_g}=\frac{\int d^4x~\sqrt{-g}\left[\frac{M_{\text P}^2}{2}R-\frac{1}{2}g^{\mu \nu}\partial_{\mu}\phi\partial_{\nu}\phi -V(\phi) \right]}{\int d^4x~\sqrt{-g}}.
\end{equation}
This modifies the Einstein's equations to
\begin{equation}
M_{\rm P}^2G_{\mu \nu}=T_{\mu \nu}-\bar{S}g_{\mu \nu},
\end{equation}
with
\begin{equation}\label{Tmunu}
T_{\mu \nu}=\partial_{\mu}\phi \partial_{\nu}\phi -g_{\mu \nu}\left[ \frac{1}{2}g^{\alpha \beta}\partial_{\alpha}\phi\partial_{\beta}\phi +V(\phi)   \right] .
\end{equation}
The Friedmann equations become
\begin{gather}
3M_{\text P}^2 H^2=\rho_{\phi}+\bar{S}, \label{f1} \\
3M_{\text P}^2(\dot{H}+H^2)=-\frac{\rho_{\phi}+3p_{\phi}}{2}+\bar{S} \label{f2},
\end{gather}
with the definitions
\begin{equation}
\rho_{\phi}=\frac{1}{2}\dot{\phi}^2+V(\phi),\quad p_{\phi}=\frac{1}{2}\dot{\phi}^2-V(\phi),\quad H=\frac{\dot{a}}{a}, \quad w_{\phi}=\frac{p_{\phi}}{\rho_{\phi}}.
\end{equation}
The $\phi$ equation of motion is still
\begin{equation}
\ddot{\phi}+3H\dot{\phi}+V'(\phi)=0.
\end{equation}

Compared to the familiar theory described by just $S$, there are now new terms $\propto \bar{S}$ in the Friedmann equations. However, using (\ref{f1}) and (\ref{f2}),
\begin{equation}
\bar{S}=\frac{\int d^3xdt~a^3\left[3M_{\text P}^2(\dot{H}+2H^2)+p_{\phi} \right]}{\int d^3xdt~a^3}=\frac{\int d^3xdt~a^3\left[ 2 \bar{S}+\frac{\rho_{\phi}-p_{\phi}}{2} \right]}{\int d^3xdt~a^3},
\end{equation}
so
\begin{equation}\label{sbar1}
\bar{S}=-\frac{\int dt~a^3V(\phi(t))}{\int dt~a^3}=-\lim_{T\rightarrow \infty} \frac{\int_{t_i}^T dt~a^3 \left[ V(\phi(t)) -V_{\min}\right] }{\int_{t_i}^T dt~a^3}-V_{\min},
\end{equation}
where $t_i$ is the time when $\phi$ becomes dynamical and $V_{\min}$ is the global minimum of the potential $V(\phi)$ that $\phi$ rolls towards. We write $\bar{S}$ as such so that the integrand of the numerator in (\ref{sbar1}) is zero at $t=T$.

It is not hard to see that the first term in $\bar{S}$ written this way vanishes on reasonable potentials and rolling $\phi$ profiles that satisfy the minimal requirements of smoothness, increasing the scale factor and decaying energy density. Explicitly, note that $\int dt~a^3V(\phi)$ is bounded by $\int dt~a^3 \rho_{\phi}$ from above; let us consider the quantity
\begin{equation}
I_1=\frac{\int_{t_i}^T dt~a^3 \rho_{\phi}}{\int_{t_i}^T dt~a^3}.
\end{equation}
As $a(t)$ increases with time, we only need to check the late-time behavior of $\rho_{\phi}$. In a typical quintessence/inflation scenario, we have $V_{\min}=0$ and $\rho_{\phi}\propto a^{-n}$ with $n(t)=\frac{3}{1+w_{\phi}(t)}>0$ throughout the evolution. Thus, in the $T\rightarrow \infty$ limit, $I_1\rightarrow 0$. This establishes $\bar{S}=0$, implying that any such scenario remains untouched in the volume normalized framework. One can also freely include ordinary matter ($\rho \propto a^{-3}$) or radiation ($\rho \propto a^{-4}$) in the scenario, because they do not contribute to $\bar{S}$, by the same argument.

Note that a constant shift of $V(\phi)$ does not change the Friedmann equations. A straightforward generalization of the analysis above shows that it is only the ``non-constant" part of $\rho_{\phi}$, $V-V_{\min}$, that drives cosmic expansion. While this follows intuitively from the CC-freedom of the volume normalized action, it can be seen nontrivially in (\ref{sbar1}); because $V(\phi(T))-V_{\min}=0$ by definition and the integrals are dominated by late-time contributions, the first term of (\ref{sbar1}) vanishes for any sensible $V(\phi)$ and $\phi$ profiles, and the sole effect of $\bar{S}$ is to neutralize a CC.

As an example, consider the ultralight pseudo-Nambu-Goldstone boson (PNGB) model for dark energy \cite{pngb}, in which the PNGB scalar field has the effective potential
\begin{equation}\label{p1}
V(\phi)=M^4\left[1- \cos\left( \frac{\phi}{f}\right)  \right].
\end{equation}
The structure of this potential is protected by symmetries in a theory from which such a potential 
is presumed to be induced in analogy with QCD, so the smallness of $m_{\phi} \sim \frac{M^2}{f} \sim 10^{-29}-10^{-34}~\text{eV}$ is natural from the particle physics point of view ($M \sim $ light neutrino mass $\sim 10^{-2}-10^{-3}~\text{eV}$, $f \sim$ global symmetry breaking scale $\sim 10^{16}-10^{19}~\text{GeV}$). At early times, $\rho_{\phi} \simeq -p_{\phi} \sim M^4$ mimics a CC; at late times, $\rho_{\phi} \sim a^{-3}$ looks like ordinary matter \cite{pngb}. This clearly guarantees $\bar{S}=0$, and thus represents a viable and theoretically favorable dark energy scenario in the volume normalized framework. The same can be said about the natural inflation model, where the QCD axion PNGB has the same potential (\ref{p1}), except that it probes the GUT scale, $M \sim 10^{16}$ GeV and $f \gsim 10^{19}$ GeV \cite{ni}. 

As an aside, we also briefly look at a minimally coupled phantom scalar \cite{psca}, which has the wrong sign kinetic term. For a homogeneous phantom scalar on flat FRW background, $\rho_{\phi}=-\frac{1}{2}\dot{\phi}^2+V(\phi)$, $p_{\phi}=-\frac{1}{2}\dot{\phi}^2-V(\phi)$, $w_{\phi}<-1$, and at late times, $\rho_{\phi}\propto a^n$ with $n>0$. The same argument above gives
\begin{equation}
-\bar{S}=\frac{\int dt~a^3 V(\phi(t))}{\int dt~a^3}>\frac{\int dt~a^3 \rho_{\phi}}{\int dt~a^3}=\infty,
\end{equation}
if one uses a $\phi$ profile that leads to cosmic expansion. This clearly makes no sense in view of (\ref{f1}), so phantom energy is ruled out in the volume normalized framework.

\section{Non-minimally Coupled Scalar} \label{nmcs}
A scalar field non-minimally coupled to gravity can be described in the volume normalized framework by
\begin{equation}\label{nm1}
\bar{S}=\frac{S}{V_g}=\frac{\int d^4x~\sqrt{-g}\left[\frac{M_0^2+f(\phi)}{2}R+\mathcal{L}_{\rm m}(g^{\mu \nu},\phi,\psi_I) \right]}{\int d^4x~\sqrt{-g}},
\end{equation}
where $M_0$ is some mass parameter, and $\mathcal{L}_{\rm m}$ is the matter Lagrangian that includes $\phi$ and the (other) Standard Model (SM) fields $\psi_I$. In principle, $M_0$ could be set to zero, under which the vacuum expectation value (VEV) of $f(\phi)$ then induces gravity; here we only make the general identification, $M_{\rm P}^2=M_0^2+\langle f(\phi) \rangle_{\rm VEV}$. In this convention, it is the metric (and its inverse) $g_{\mu \nu}$ that all the fields in $\mathcal{L}_{\rm m}$ couple to; for example,
\begin{equation}
\mathcal{L}_{\rm m}(g^{\mu \nu},\phi,\psi) \supset -\frac{1}{2}g^{\mu \nu} \partial_{\mu}\phi \partial_{\nu} \phi -V(\phi).
\end{equation}
We emphasize this point because it makes the form (\ref{nm1}) unique if we are to retain the solution of the CC problem. In general, because of the conformal factor in the numerator, one could introduce scalar field dependence in the denominator in (\ref{nm1}) when defining a scalar-tensor theory in the volume normalized framework of (\ref{ac2}).\footnote{We thank the referee for bringing up this point.} However, the matter quantum loops will generate a CC term in the numerator, $\int d^4x\sqrt{-g} \Lambda_{\rm m}$, with $\Lambda_{\rm m}$ being a veritable constant independent of the fields. In order for this quantum correction to CC to drop out of the physical picture via (\ref{ac2}),
one really needs to keep the volume factor in the denominator 
independent of the scalar field; this selects the form (\ref{nm1}). We also call the fields written in (\ref{nm1}) in the ``Jordan frame," to relate to the well known nomenclature.    

Then the Einstein's equations become
\begin{equation}
\left[ M_0^2+f(\phi) \right]G_{\mu \nu}=T_{\mu \nu}-\bar{S}g_{\mu \nu}+\left( \nabla_{\mu}\nabla_{\nu}-g_{\mu \nu}\square \right)f(\phi),
\end{equation}
where $T_{\mu \nu}$ is defined as usual,
\begin{equation}
T_{\mu \nu}=\frac{-2}{\sqrt{-g}}\frac{\partial(\sqrt{-g}\mathcal{L}_{\rm m}(g^{\mu \nu},\phi,\psi))}{\partial g^{\mu \nu}}.
\end{equation}
Using only $\mathcal{L}_{\rm m}(g^{\mu \nu},\phi,\psi) \rightarrow -\frac{1}{2}g^{\mu \nu} \partial_{\mu}\phi \partial_{\nu} \phi -V(\phi)$, we continue to write the Friedmann equations (with the replacement $M_{\rm P}\rightarrow M_0$) in the form of (\ref{f1}) and (\ref{f2}), but now with 
\begin{gather}
\rho_{\phi} = \frac{1}{2}\dot{\phi}^2+V(\phi)-3Hf'(\phi)\dot{\phi},\\
p_{\phi}=\frac{1}{2}\dot{\phi}^2-V(\phi)+ 2Hf'(\phi) \dot{\phi}+f'(\phi)\ddot{\phi}+f''(\phi)\dot{\phi}^2,
\end{gather}
and the $\phi$ equation is
\begin{equation}\label{phi2}
\ddot{\phi}+3H\dot{\phi}+V'(\phi)-3(\dot{H}+2H^2)f'(\phi)=0.
\end{equation}
Plugging the Friedmann equations back into (\ref{nm1}),
\begin{align}
\bar{S}&=\frac{1}{\int dt~a^3}\int dt~a^3 \left\{  -V(\phi)+ \frac{3}{2} \left[ 3Hf'(\phi) \dot{\phi}+f'(\phi)\ddot{\phi}+f''(\phi)\dot{\phi}^2 \right] \right\} \label{line1}\\
&=-\frac{1}{\int dt~a^3}\int dt~a^3 \left\{ \rho_{\phi} -\frac{1}{2}\dot{\phi}^2- \frac{3}{2} \left[ Hf'(\phi) \dot{\phi}+f'(\phi)\ddot{\phi}+f''(\phi)\dot{\phi}^2 \right] \right\}.  \label{line2}
\end{align}
Compared to (\ref{sbar1}), the new terms in the integrand of (\ref{line1}) are all proportional to the time derivatives of $\phi$, which vanish at the end of the cosmic evolution. Then we have $\bar{S}=-V_{\min}$ by the same argument in the last section for reasonable potentials and $\phi$ profiles. Also, note that by (\ref{line2}), for $\rho_{\phi}$ proportional to the negative powers of $a(t)$, which necessarily means $V_{\min}=0$, we have $\bar{S}=0$ identically. Therefore, the classical pictures of cosmic expansion in the non-minimally coupled models are exactly the same in the frameworks of $S$ and $\bar{S}$.

This result applies generically to a wide range of non-minimally coupled dark energy \cite{nmcintro,nmcintro2,nmc21,nmc22,nmc223,nmc23,nmc24} and inflation models \cite{infnmc01,infnmc02,infnmc03,infnmc21,infnmc22,infnmc23,infnmc24}. Notably, the Higgs inflation model \cite{higgs1,higgs2} uses the radial mode of the Higgs doublet as the non-minimally coupled scalar and needs no extra field beyond the SM. There the equations of motion receive extra contributions from the coupling between Higgs and other SM fields, but these local effects clearly do not affect the argument above.

So far we have worked in the Jordan frame. In the non-minimally coupled scalar setup, it is also popular to transform to the Einstein frame, in which we will demonstrate the result above with the choice \cite{infnmc21,infnmc22,infnmc23,infnmc24,higgs1,higgs2}
\begin{equation}
f(\phi)= \xi \phi^2, \quad V(\phi)=\frac{\lambda}{4}\left( \phi^2-v^2 \right)^2.
\end{equation}
For concreteness, we take $\phi$ to be the radial Higgs mode, and then $M_0 \approx M_{\rm P}$ with very high accuracy. Following \cite{higgs1,higgs2}, the field redefinitions read
\begin{equation}
g_{\mu \nu}=\Omega^{-2}\hat{g}_{\mu \nu},\quad \Omega^2=1+\frac{\xi \phi^2}{M_{\rm P}^2},\quad \quad \frac{d\chi}{d\phi}=\sqrt{\frac{\Omega^2+\frac{6\xi^2\phi^2}{M_{\rm P}^2}}{\Omega^4}},
\end{equation}
upon which (\ref{nm1}) with $\mathcal{L}_{\rm m}(g^{\mu \nu},\phi,\psi) \rightarrow -\frac{1}{2}g^{\mu \nu} \partial_{\mu}\phi \partial_{\nu} \phi -V(\phi)$ becomes
\begin{equation}\label{nmcs2}
\bar{S}=\frac{\int d^4x~\sqrt{-\hat{g}}\left[\frac{M_{\text P}^2}{2}\hat{R}-\frac{1}{2}\hat{g}^{\mu \nu}\partial_{\mu}\chi\partial_{\nu}\chi -U(\chi) \right]}{\int d^4x~\Omega^{-4}(\chi)\sqrt{-\hat{g}}},
\end{equation}
where $\hat{R}$ is calculated from $\hat{g}_{\mu \nu}$, and
\begin{equation}
U(\chi)=\Omega^{-4}(\chi)\frac{\lambda}{4} \left( \phi^2(\chi)-v^2 \right)^2.
\end{equation}
On rolling solutions that lead to cosmic expansion, the field values are large at early times and small at the end of the evolution, so $\Omega^{-4}$ gradually increases to 1. Then the situation is essentially reduced to that of a minimally coupled scalar $\chi$ with $U_{\min}=0$, because the integrals in (\ref{nmcs2}) are clearly dominated by late-time contributions. Repeating the analysis in the last section step by step then yields $\bar{S}=0$, as expected.

\section{Chaplygin Gas}\label{cg}
Chaplygin gas is an exotic cosmic fluid that has the equation of state $p=-\frac{A}{\rho}$. In a flat FRW universe, energy conservation implies that
\begin{equation}\label{chap1}
\rho = \sqrt{A+\frac{B}{a^6}},
\end{equation}
which represents a unified dark matter/energy fluid that transitions from a matter, $\rho \simeq \sqrt {B} /a^3$
(and a negligible pressure),  to a CC,
$\rho \simeq -p \simeq \sqrt{A}$ \cite{chap}. Microscopically, it has been shown that the Chaplygin gas equation of state can be induced by a 3-brane living in a 5D Minkowski space on the 4D braneworld transverse to its lightcone \cite{microchap-1,microchap0,microchap1,microchap2,microchap3}; it remains to be checked if this picture can be made compatible with (\ref{pr1}). Anyway, we present the Chaplygin gas scenario as a negative example.

The profile (\ref{chap1}) can be realized by a scalar field minimally coupled to gravity with the potential
\begin{equation}
V(\phi)=\frac{\sqrt{A}}{2}\left[ \text{cosh}\left( \frac{\sqrt{3}\phi}{M_{\rm P}}\right) +\frac{1}{ \text{cosh}\left( \frac{\sqrt{3}\phi}{M_{\rm P}}\right) } \right].
\end{equation}
However, in the volume normalized framework, the analysis in section \ref{mcs} shows that $V_{\min}=\sqrt{A}$ does not contribute to the Friedmann equations. The effective potential in this case is
\begin{equation}
V_{\rm eff}=V(\phi)-V_{\min}=\frac{\sqrt{A}}{2}\left[ \text{cosh}\left( \frac{\sqrt{3}\phi}{M_{\rm P}}\right) +\frac{1}{ \text{cosh}\left( \frac{\sqrt{3}\phi}{M_{\rm P}}\right) } -2 \right],
\end{equation}
so the Chaplygin gas solution is no longer valid. In fact,
\begin{equation}
V_{\rm eff}\propto \frac{ \left[ \text{cosh}\left( \frac{\sqrt{3}\phi}{M_{\rm P}}\right)-1\right]^2}{\text{cosh}\left( \frac{\sqrt{3}\phi}{M_{\rm P}}\right)},
\end{equation}
which describes just another quintessence scenario interpolating between known models, $V(\phi) \propto \left[ \text{cosh}\left( \frac{\sqrt{3}\phi}{M_{\rm P}}\right)-1\right]^q$ with $q=1,2$ \cite{quint1}. The same argument here also applies to generalized Chaplygin gas models \cite{gchap}, simply because the extra terms in the Friedmann equations nullify any CC-like component.

\section{Fluctuations and Quantum Effects in Inflation}\label{conc}
We first make some general observations about the volume normalized action, (\ref{ac2}). Its variational procedure yields
\begin{equation}
\delta \bar{S}=\delta\left( \frac{S}{V_g}\right)=\frac{1}{V_g}\left(\delta S-\frac{S}{V_g}\delta
V_g \right),
\end{equation}
so all non-gravitational local physics encoded in $S$ are left unchanged, while the Einstein's equations are modified by the extra term $\propto \bar{S}$. The latter is then the key to understand the differences between the theories described by $S$ and $\bar{S}$. In both dark energy and inflationary scenarios,  the scale factor $a(t)$ increases with time and becomes very large at the end of the evolution, so $S$ is dominated by the late-time behavior of its integrand, which implies that, $\bar{S}=-V_{\min}$. Without surprise, this is just a roundabout way to see that the volume normalized action automatically gets rid of any CC-like contribution to the cosmic equation of state. 

We then move on to field fluctuations. For the scalar fluctuations $\delta \phi(\vec{x},t)$, both classical and quantum, they are merely local effects and do not contribute to $\bar{S}$. The same can be said about the metric fluctuations $\delta  g_{\mu \nu}(\vec{x},t)$. Finite temperature has also no effect on $\bar{S}$, which depends only on the late-time picture of the Universe that approaches a zero temperature state. Thus, the dynamics of field fluctuations do not differ from that obtained via $S$, except that $\delta  g_{\mu \nu}(\vec{x},t)$ are effectively classical, owing to the vanishing Planck constant for gravity loops, implied by the full action (\ref{pr1}). We will see below the nontrivial effects of this fact in the inflationary cosmology.

Because the inflaton is still quantized with $\hbar$, the inflationary predictions for primordial scalar modes do not change in the new framework. This can be seen in two ways. First, even though the scalar perturbation is a mixture of the inflaton and the metric fluctuation, it is  possible to choose a gauge in which the scalar modes are due entirely to the 
inflaton fluctuations. Second, one needs not choose a gauge, but rather organize the scalar perturbations into gauge invariant quantities, which are to be diagonalized and canonically normalized before quantization. For example, during the de Sitter stage of a minimally coupled model, the quadratic action for scalar perturbations can be reduced to \cite{infreview1,infreview2,infreview3,infreview4,infreview5}
\begin{equation}\label{muk}
S_2=\frac{1}{2}\int d^3xd\eta\left[ (u')^2-(\partial_iu)^2+\frac{z''}{z}u^2 \right],
\end{equation} 
where $\eta$ is the conformal time and primes are derivatives with respect to it. The gauge invariant quantities are
\begin{equation}
u=z\mathcal{R},\quad \mathcal{R}=\left( \Psi+\frac{H}{\dot{\phi}}\delta \phi \right), \quad z^2=\frac{a^2\dot{\phi}^2}{H^2},
\end{equation}
where $\mathcal{R}$ is known as the comoving curvature perturbation, and $\Psi$ is defined in the way that the scalar and tensor metric fluctuations are expressed as
\begin{equation}\label{fluc}
ds^2=-(1+2\Phi)dt^2+2a\partial_iBdx^idt+a^2\left[ (1-2\Psi)\delta_{ij}+ 2\partial_i \partial_j E+h_{ij} \right]dx^idx^j.
\end{equation} 
One can then go to a gauge with $\Psi=0$, so that $u$ is due to $\delta \phi$ only. One can also directly quantize $u$ from (\ref{muk}), without fixing $\Psi$ or $\delta \phi$. In the latter case, the quantized scalar modes are completely equivalent regardless of whether the scalar metric fluctuations are fundamentally classical or quantum, thus leading to the same predicted power spectra. The subtleties of diagonalizing a quantum field and a classical field are addressed in Appendix \ref{2scalar}.

On the other hand, the tensor metric fluctuations $h_{ij}$ in (\ref{fluc}), which satisfy the transverse traceless 
condition $\partial^ih_{ij}=h^i_i=0$,  and evolve separately, are not quantized in the volume normalized 
framework, thus giving vanishing quantum correlation functions and power spectra. This implies that no primordial 
tensor modes arise during inflation -- that is, the theory predicts a zero tensor-to-scalar ratio, $r=0$, irrespective of the inflationary model. Thus,  the framework (\ref{pr1}) with $\hbar V_g/V_F\to 0$ could be falsified  
by observational data that would deny this null prediction, for example, by the detection of the primordial 
gravitational waves.

\section*{Acknowledgements}
We thank Justin Khoury  and Archil Kobakhidze for useful discussions. 
Both GG and SY are supported by NASA grant NNX12AF86G S06 and NSF grant
PHY-1316452. SY is also supported by the NYU James Arthur graduate fellowship. 
GG acknowledges a membership at the NYU-ECNU Joint Physics Research 
Institute in Shanghai.  

\appendix

\section{A Tale of Two Scalars} \label{2scalar}
Here we use a toy model to illustrate the subtleties of diagonalizing a quantum field and a classical field. Consider the following theory of two coupled scalar fields in a Minkowski space,
\begin{equation}\label{toy1}
S_{\rm toy}=\int d^4x\left[-\frac{1}{2}(\partial \sigma)^2-\frac{1}{2}(\partial \tau)^2 -\frac{1}{2}m^2(\sigma^2+\tau^2)-b^2 \sigma \tau  \right],
\end{equation}
where $b^2 \leq m^2$ is a coupling constant. We demand that $\sigma$ is a quantum field but $\tau$ is a classical field, 
being analogs of  the inflaton and metric perturbations, respectively. The fields have commutators
\begin{equation}\label{comm1}
[\sigma(\vec{x},t),\dot{\sigma}(\vec{y},t)]=i\hbar\delta(\vec{x}-\vec{y}), \quad [\tau(\vec{x},t),\dot{\tau}(\vec{y},t)]= 0,
\end{equation}
and all others are zero. 
The usual procedure of diagonalization implements the field redefinitions
\begin{equation}
\Lambda=\sigma+\tau,\quad \Theta= \sigma-\tau,
\end{equation}
upon which
\begin{equation}
S_{\rm toy}=\int d^4x\left[-\frac{1}{4}(\partial \Lambda)^2-\frac{1}{4}(\partial \Theta)^2-\frac{1}{4}(m^2+b^2)\Lambda^2-\frac{1}{4}(m^2-b^2)\Theta^2  \right].
\end{equation}
Naively, if we were to quantize both $\Lambda$ and $\Theta$ (because they each appear to contain a quantum component), it would seem that we are getting two quantum fields out of one quantized  field and one classical field, which cannot be consistent. 
However, (\ref{comm1}) implies that there are now nontrivial cross terms among the new commutators,
\begin{equation}
[\Lambda(\vec{x},t),\dot{\Theta}(\vec{y},t)]=[\Theta(\vec{x},t),\dot{\Lambda}(\vec{y},t)]=i\hbar\delta(\vec{x}-\vec{y}).
\end{equation}
Therefore, $\Lambda$ and $\Theta$ are not independent quantum degrees of freedom, in the sense that a specific combination of them is classical, in this case $\tau=\frac{\Lambda-\Theta}{2}$. Then the ``undiagonalized" form (\ref{toy1}) is actually more suitable for the computation of quantum correlation functions in the standard language of quantum field theory, with $\tau$ treated as a classical source. In the context of inflationary scalar perturbations above, this means that we need not obtain the diagonalized form (\ref{muk}), but just compute the correlation functions of $\delta \phi$ in any gauge to get the correct power spectra.

\end{document}